\documentclass[conference]{IEEEtran}

\usepackage[pdftex]{graphicx}
\usepackage[cmex10]{amsmath}
\usepackage{array}
\usepackage{listings}
\usepackage{multicol}
\usepackage{subfig}
\usepackage{enumitem}
\usepackage{caption}

\begin{document}

\title{\textbf{Parallel Simulations for Analysing Portfolios of Catastrophic Event Risk}}

\author{\IEEEauthorblockN{Aman Bahl}
\IEEEauthorblockA{Center for Security, Theory and Algorithmic Research\\International Institute of Information Technology\\Hyderabad, India\\
Email: aman.kumar@research.iiit.ac.in}
\and
\IEEEauthorblockN{Oliver Baltzer, Andrew Rau-Chaplin and Blesson Varghese}
\IEEEauthorblockA{Risk Analytics Laboratory\\Dalhousie University\\Halifax, Canada\\
Email: \{obaltzer, arc, varghese\}@cs.dal.ca}}

\maketitle

\begin{abstract}
At the heart of the analytical pipeline of a modern quantitative insurance/reinsurance company is a stochastic simulation technique for portfolio risk analysis and pricing process referred to as Aggregate Analysis. Support for the computation of risk measures including Probable Maximum Loss (PML) and the Tail Value at Risk (TVAR) for a variety of types of complex property catastrophe insurance contracts including Cat eXcess of Loss (XL), or Per-Occurrence XL, and Aggregate XL, and contracts that combine these measures is obtained in Aggregate Analysis.

In this paper, we explore parallel methods for aggregate risk analysis. A parallel aggregate risk analysis algorithm and an engine based on the algorithm is proposed. This engine is implemented in C and OpenMP for multi-core CPUs and in C and CUDA for many-core GPUs. Performance analysis of the algorithm indicates that GPUs offer an alternative HPC solution for aggregate risk analysis that is cost effective. The optimised algorithm on the GPU  performs a 1 million trial aggregate simulation with 1000 catastrophic events per trial on a typical exposure set and contract structure in just over 20 seconds which is approximately 15x times faster than the sequential counterpart. This can sufficiently support the real-time pricing scenario in which an underwriter analyses different contractual terms and pricing while discussing a deal with a client over the phone.
\end{abstract}


\IEEEpeerreviewmaketitle

\section{Introduction}

Risk analytics, the model based computational analysis of risk \cite{1}, has become an integral part of business processes in domains ranging from financial services to engineering. Simulation based risk analytics has been applied to areas as diverse as analysis of catastrophic events \cite{16,18}, financial instruments \cite{19}, structures \cite{3}, chemicals \cite{4}, diseases \cite{17}, power systems \cite{6}, nuclear power plants \cite{15}, radioactive waste disposal \cite{13} and terrorism \cite{14}. In many of these areas models must both consume huge data sets and perform hundreds of thousands or even millions of simulations making the application of parallel computing techniques very attractive. For financial analysis problems, especially those concerned with the pricing of assets, parallelism and high-performance computing has been applied to very good effect (for example \cite{20, 21, 22, 23} and \cite{24}). However, in the insurance and reinsurance settings, where data sizes are arguably as large or larger, relatively fewer HPC based methods have been reported on.

In the insurance and reinsurance settings companies hold portfolios of contracts that cover risks associated with catastrophic events such as earthquakes, hurricanes and floods. In order to have a marketplace for such risk it is critical to be able to efficiently quantify individual risks and portfolios of risks. The analytical pipeline of the modern quantitative insurance or reinsurance company typically consists of three major stages: (i) risk assessment, (ii) portfolio risk management and pricing, and (iii) enterprise risk management.

In the first stage, catastrophe models \cite{25} are used to provide scientifically credible loss estimates for individual risks by taking two inputs. Firstly, stochastic event catalogs which are a mathematical representation of the natural occurrence patterns and characteristics of catastrophe perils such as hurricanes, tornadoes, severe winter storms or earthquakes. Secondly, exposure databases that describe thousands or millions of buildings to be analysed, their construction types, location, value, use, and coverage. Each event-exposure pair is then analysed by a risk model that quantifies the hazard intensity at the exposure site, the vulnerability of the building and resulting damage level, and the resultant expected loss, given the customer's financial terms. The output of a catastrophe model is an Event Loss Table (ELT) which specifies the probability of occurrence and the expected loss for every event in the catalog. However, an ELT does not capture which events are likely to occur in a contractual year, in which order, and how they will interact with complex treaty terms to produce an aggregated loss.

Reinsurers may have thousands or tens of thousands of contracts and must analyse the risk associated with their whole portfolio. These contracts often have an `eXcess of Loss' (XL) structure and can take many forms, including (i) Cat XL or Per-Occurrence XL contracts providing coverage for single event occurrences up to a specified limit with an optional retention by the insured and (ii) Aggregate XL contracts (also called stop-loss contracts) providing coverage for multiple event occurrences up to a specified aggregate limit and with an optional retention by the insured. In addition, combinations of such contract terms providing both Per-Occurrence and Aggregate features are common.

In the second stage of the analysis pipeline, portfolio risk management and pricing of portfolios of contracts necessitates a further level of stochastic simulation, called aggregate analysis \cite{s1, s2, s3, s4, s5, s6} (see Figure~\ref{figure1}). Aggregate analysis is a form of Monte Carlo simulation in which each simulation trial represents an alternative view of which events occur and in which order they occur within a predetermined period, i.e., a contractual year. In order to provide actuaries and decision makers with a consistent lens through which to view results, rather than using random values generated on-the-fly, a pre-simulated Year Event Table (YET) containing between several thousand and millions of alternative views of a single contractual year is used. The output of aggregate analysis is a Year Loss Table (YLT). From a YLT, a reinsurer can derive important portfolio risk metrics such as the Probable Maximum Loss (PML) and the Tail Value at Risk (TVAR) which are used for both internal risk management and reporting to regulators and rating agencies. Furthermore, these metrics then flow into the final stage in the risk analysis pipeline, namely Enterprise Risk Management, where liability, asset, and other forms of risks are combined and correlated to generate an enterprise wide view of risk.

In this paper, we explore parallel methods for aggregate risk analysis. A parallel aggregate risk analysis algorithm and an engine based on the algorithm is proposed. This engine is implemented in C and OpenMP for multi-core CPU platforms and in C and CUDA for many-core GPU platforms. Experimental studies are pursued on both the platforms.

GPUs offer an alternative machine architecture in three ways. GPUs provide, firstly, lots of cycles for independent parallelism, secondly, fast memory access under the right circumstances, and finally, fast mathematical computation at relatively low costs. A wide range of applications in the scientific and financial domains, for example those presented in \cite{40, 41, 42, 43, 44} and \cite{45}, benefit from the merits of the GPU architecture. Our current aggregate analysis engine takes full advantage of the high levels of parallelism, some advantage of the fast shared memory access, but relatively little advantage of the GPUs fast numerical performance. Overall, our experiments suggest that GPUs offer a cost effective HPC architecture for aggregate risk analysis.

The remainder of this paper is organised as follows. Section \ref{aggregateriskanalysis} describes the mathematical notations of the aggregate risk analysis algorithm. Section \ref{experimentalstudies} describes the experimental evaluation of the implementation of the algorithm on the CPU and GPU and factors affecting the performance of the algorithm. Section \ref{conclusion} concludes the paper and summarises the performance results.

\section{Aggregate Risk Analysis}
\label{aggregateriskanalysis}

\begin{figure}
	\centering
	\includegraphics[width = 0.5\textwidth]{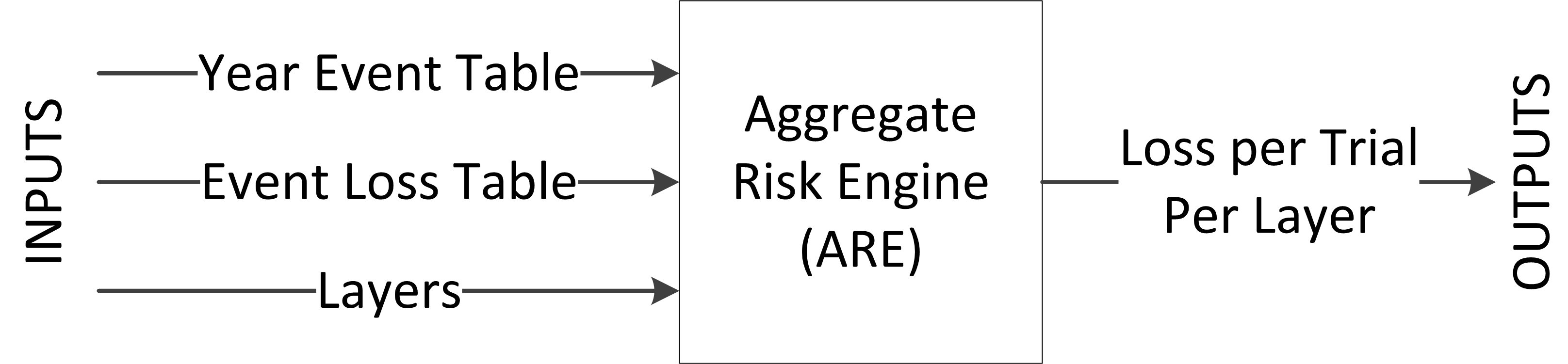}
	\caption{The inputs and output of the Aggregate Risk Engine (ARE)}
	\label{figure1}
\end{figure}

The \textbf{A}ggregate \textbf{R}isk \textbf{E}ngine, referred to as ARE (see Figure \ref{figure1}) is considered in this section. The description of ARE is separated out as the inputs to the engine, the basic sequential algorithm for aggregate analysis and the output of the engine.

\subsection{Inputs}
Inputs to the Aggregate Risk Engine are three main components:

\begin{enumerate}
\item The Year Event Table (YET), denoted as $YET$, is a database of pre-simulated occurrences of events from a catalog of stochastic events. Each record in a YET called a ``trial'', denoted as $T_i$, represents a possible sequence of event occurrences for any given year. The sequence of events is defined by an ordered set of tuples containing the ID of an event and the time-stamp of its occurrence in that trial $T_i = \{(E_{i, 1}, t_{i, 1}), \dots, (E_{i, k}, t_{i, k})\}$. The set is ordered by ascending time-stamp values. A typical YET may comprise thousands to millions of trials, and each trial may have approximately between 800 to 1500 `event time-stamp' pairs, based on a global event catalog covering multiple perils. The YET can be represented as:

\begin{equation*}
YET=\left\{ T_i = \left\{(E_{i, 1}, t_{i, 1}), \dots, (E_{i, k}, t_{i, k})\right\} \right\}
\end{equation*}
\begin{center}
where $i = 1, 2, \dots$ and $k = 1, 2, \dots, 800-1500$.
\end{center}

\item Event Loss Tables, denoted as $ELT$, represent collections of specific events and their corresponding losses with respect to an exposure set. An event may be part of multiple ELTs and associated with a different loss in each ELT. For example, one ELT may contain losses derived from one exposure set while another ELT may contain the same events but different losses derived from a different exposure set. Each ELT is characterised by its own metadata including information about currency exchange rates and terms that are applied at the level of each individual event loss. Each record in an ELT is denoted as event loss $EL_{i} = \{E_{i}, l_{i}\}$, and the financial terms associated with the ELT are represented as a tuple $\mathcal{I} = (\mathcal{I}_{1}, \mathcal{I}_{2}, \dots)$. A typical aggregate analysis may comprise 10,000 ELTs, each containing 10,000-30,000 event losses with exceptions even up to 2,000,000 event losses. The ELTs can be represented as:

\begin{equation*}
ELT=\left\{
	\begin{array}{c}
	EL_{i} = \{E_{i}, l_{i}\}, \\
	\mathcal{I} = (\mathcal{I}_{1}, \mathcal{I}_{2}, \dots)
	\end{array}\right\}
\end{equation*}
\begin{center}
with $i = 1, 2, \dots , 10,000-30,000$.
\end{center}

\item Layers, denoted as $L$, cover a collection of ELTs under a set of layer terms. A single layer $L_i$ is composed of two attributes. Firstly, the set of ELTs $\mathcal{E} = \{ELT_1, ELT_2, \dots, ELT_j\}$, and secondly, the Layer Terms, denoted as $\mathcal{T} = (\mathcal{T}_{OccR}, \mathcal{T}_{OccL}, \mathcal{T}_{AggR}, \mathcal{T}_{AggL})$. A typical layer covers approximately 3 to 30 individual ELTs. The Layer can be represented as:

\begin{equation*}
L=\left\{
	\begin{array}{c}
	\mathcal{E} = \{ELT_1, ELT_2, \dots, ELT_j\}, \\
	\mathcal{T} = (\mathcal{T}_{OccR}, \mathcal{T}_{OccL}, \mathcal{T}_{AggR}, \mathcal{T}_{AggL})
	\end{array}\right\}
\end{equation*}
\begin{center}
with $j = 1, 2, \dots, 3-30$.
\end{center}

\end{enumerate}

\subsection{Algorithm Sketch}

The principal sequential algorithm for aggregate analysis utilised in ARE consists of two stages: 1) a preprocessing stage in which data is loaded into local memory, and 2) the analysis stage performing the simulation and producing the resulting YLT output. The algorithm for the analysis stage is as follows:

\begin{table}[ht]
	\caption*{}
	\vspace{-0.5cm}
	\begin{center}
		\begin{tabular}{ m{0.46\textwidth} }
			\hline
			\normalsize{\textit{\textbf{Basic Algorithm for Aggregate Risk Analysis}}}\\
			\hline
			\hline
			\begin{enumerate} [leftmargin = 0cm]
				\normalsize{
				\item[1] \textbf{for} all $a \in L$
				\item[2] \hspace{0.7cm}\textbf{for} all $b \in YET$
				\item[3] \hspace{1.4cm}\textbf{for} all $c \in (EL \in a)$
				\item[4] \hspace{2.1cm}\textbf{for} all $d \in (Et \in b)$
				\item[5] \hspace{2.8cm}$x_{d} \leftarrow E \in d$ in $El \in f$,
				\item[]	 \hspace{2.8cm}where $f \in ELT$ and $(EL \in f) = c$
				\item[6] \hspace{2.1cm}\textbf{for} all $d \in (Et \in b)$
				\item[7] \hspace{2.8cm}$l_{x_{d}} \leftarrow$ Apply Financial Terms$(\mathcal{I})$
				\item[8] \hspace{2.1cm}\textbf{for} all $d \in (Et \in b)$
				\item[9] \hspace{2.8cm}$lo_{x_{d}} += l_{x_{d}}$
				\item[10] \hspace{1.4cm}\textbf{for} all $d \in (Et \in b)$
				\item[11] \hspace{2.1cm}$lo_{x_{d}} = min(max(lo_{x_{d}} - \mathcal{T}_{OccR}, 0), \mathcal{T}_{OccL})$
				\item[12] \hspace{1.4cm}\textbf{for} all $d \in (Et \in b)$
				\item[13] \hspace{2.1cm}$lo_{x_{d}} = \sum\limits_{i=1}^{d}lo_{x_{i}}$
				\item[14] \hspace{1.4cm}\textbf{for} all $d \in (Et \in b)$
				\item[15] \hspace{2.1cm}$lo_{x_{d}} = min(max(lo_{x_{d}} - \mathcal{T}_{AggR}, 0), \mathcal{T}_{AggL})$
				\item[16] \hspace{1.4cm}\textbf{for} all $d \in (Et \in b)$
				\item[17] \hspace{2.1cm}$lo_{x_{d}} = lo_{x_{d}} - lo_{x_{d - 1}}$
				\item[18] \hspace{1.4cm}\textbf{for} all $d \in (Et \in b)$
				\item[19] \hspace{2.1cm}$lr += lo_{x_{d}}$
				}
			\end{enumerate}
			\\ \hline
		\end{tabular}
	\end{center}
\end{table}

In the preprocessing stage the input data, $YET$, $ELT$ and $L$, is loaded into memory.

The aggregate analysis stage is composed of four steps which are all executed for each Layer and each trial in the YET. In the first step (lines 3-5), for each event of a trial its corresponding event loss in the set of ELTs associated with the Layer is determined.

In the second step (lines 6 and 7), a set of financial terms is applied to each event loss pair extracted from an ELT. In other words, contractual financial terms to the benefit of the layer are applied in this step. For this the losses for a specific event's net of financial terms $\mathcal{I}$ are accumulated across all ELTs into a single event loss (lines 8 and 9).

In the third step (lines 10-13), the event loss for each event occurrence in the trial, combined across all ELTs associated with the layer, is subject to occurrence terms ($\mathcal{T}_{OccR}$, and $\mathcal{T}_{OccL}$). Occurrence terms are part of the layer terms (refer Figure~\ref{table2}) and applicable to individual event occurrences independent of any other occurrences in the trial. The occurrence terms capture the specific contractual properties of Cat XL and Per-Occurrence XL treaties as they apply to individual event occurrences only. The event losses net of occurrence terms are then accumulated into a single aggregate loss for the given trial.

In the fourth and final step (lines 14-19), the aggregate terms are applied to the trial's aggregate loss for a layer. Unlike occurrence terms, aggregate terms are applied to the cumulative sum of occurrence losses within a trial and thus the result depends on the sequence of prior events in the trial. This behaviour captures the properties of common Stop-Loss or Aggregate XL contracts. The aggregate loss net of the aggregate terms is referred to as the trial loss or the year loss and stored in a Year Loss Table (YLT) as the result of the aggregate analysis.

\begin{table*}
	\begin{center}
	\caption{Layer Terms applicable to Aggregate Risk Analysis}
	\begin{tabular}{ | c | l | p {12cm} |}
		\hline
		\hline
		Notation				&	Terms							&	Description\\ \hline
		$\mathcal{T}_{OccR}$	&	Occurrence Retention			&	Retention or deductible of the insured for an individual occurrence loss \\ \hline
		$\mathcal{T}_{OccL}$	&	Occurrence Limit				&	Limit or coverage the insurer will pay for occurrence losses in excess of the retention \\ \hline
		$\mathcal{T}_{AggR}$	&	Aggregate Retention				&	Retention or deductible of the insured for an annual cumulative loss \\ \hline
		$\mathcal{T}_{AggL}$	&	Aggregate Limit				    &	Limit or coverage the insurer will pay for annual cumulative losses in excess of the aggregate retention \\
		\hline \hline
	\end{tabular}
	\label{table2}
	\end{center}
\end{table*}

\subsection{Output}
The algorithm will provide an aggregate loss value for each trial denoted as $lr$ (line 19). Then filters (financial functions) are applied on the aggregate loss values.

\section{Experimental Evaluation}
\label{experimentalstudies}

The aims of the ARE are to handle large data, organise input data in efficient data structures, and define the granularity at which parallelism can be applied on the aggregate risk analysis problem to achieve a significant speedup. This section confirms through experimental evaluation of a number of implementations of the aggregate risk analysis algorithm as to how the aims of ARE are achieved. The hardware platforms used in the experimental evaluation are firstly considered.

\subsection{Platform}

Two hardware platforms were used in the experimental evaluation of the aggregate risk algorithm.

\subsubsection{Platform 1 - A Multi-core CPU}

The multi-core CPU employed in this evaluation was a 3.40 GHz quad-core Intel(R) Core (TM) i7-2600 processor with 16.0 GB of RAM. The processor had 256 KB L2 cache per core, 8MB L3 cache and maximum memory bandwidth of 21 GB/sec. Both sequential and parallel versions of the aggregate risk analysis algorithm were implemented on this platform. The sequential version was implemented in C++, while the parallel version was implemented in C++ and OpenMP. Both versions were compiled using the GNU Compiler Collection g++ 4.4 using the ``-O3'' and ``-m64'' flags.

\subsubsection{Platform 2 - A Many-Core GPU}

A NVIDIA Tesla C2075 GPU, consisting of 448 processor cores (organised as 14 streaming multi-processors each with 32 symmetric multi-processors) and a global memory of 5.375 GB was employed in the GPU implementations of the aggregate risk analysis algorithm. CUDA is employed for a basic GPU implementation of the aggregate risk analysis algorithm and an optimised implementation.

\subsection{Implementation}

Four variations of the Aggregate Risk Analysis algorithm are presented in this section. They are (i) a sequential implementation, (ii) a parallel implementation using OpenMP for multi-cores CPUs, (iii) a basic GPU implementation and (iv) an optimised/``chunked'' GPU implementation. In all implementations a single thread is employed per trial, $T_{id}$. The key design decision from a performance perspective is the selection of a data structure for representing Event Loss Tables (ELTs). ELTs are essentially dictionaries consisting of key-value pairs and the fundamental requirement is to support fast random key lookup. The ELTs corresponding to a layer were implemented as direct access tables. A direct access table is a highly sparse representation of a ELT, one that provides very fast lookup performance at the cost of high memory usage. For example, consider an event catalog of 2 million events and a ELT consisting of 20K events for which non-zero losses were obtained. To represent the ELT using a direct access table, an array of 2 million loss are generated in memory of which 20K are non-zero loss values and the remaining 1.98 million events are zero. So if a layer has 15 ELTs, then $15 \times 2$ million $= 30$ million event-loss pairs are generated in memory.

A direct access table was employed in all implementations over any alternate compact representation for the following reasons. A search operation is required to find an event-loss pair in a compact representation. If sequential search is adopted, then $O(n)$ memory accesses are required to find an event-loss pair. Even if sorting is performed in a pre-processing phase to facilitate a binary search, then $O(log(n))$ memory accesses are required to find an event-loss pair. If a constant-time space-efficient hashing scheme, such as cuckoo hashing \cite{29} is adopted then only a constant number of memory accesses may be required but this comes at considerable implementation and run-time performance complexity. This overhead is particularly high on GPUs with their complex memory hierarchies consisting of both global and shared memories. Compact representations therefore place a very high cost on the time taken for accessing an event-loss pair. Essentially the aggregate analysis process is memory access bound. For example, to perform aggregate analysis on a YET of 1 million trials (each trial comprising 1000 events) and for a layer covering 15 ELTs, there are $1000 \times 1$ million $ \times 15 = 15$ billion events, which requiring random access to 15 billion loss values. Direct access tables, although wasteful of memory space, allow for the fewest memory accesses as each lookup in an ELT requires only one memory access per search operation.

\subsubsection{Basic Implementations}
The data structures used for the basic implementations are:
\begin{enumerate}
\item[(i)] A vector consisting of all $E_{i,k}$ that contains approximately 800M-1500M integer values requiring 3.2GB-6GB memory.
\item[(ii)] A vector of 1M integer values indicating trial boundaries to support the above vector requiring 4MB memory.
\item[(iii)] A structure consisting of all $El_{i}$ that contains approximately 100M-300M integer and double pairs requiring 1.2GB-3.6GB.
\item[(iv)] A vector to support the above vector by providing ELT boundaries containing approximately 10K integer values requiring 40KB.
\item[(v)] A number of smaller vectors for representing $\mathcal{I}$ and $\mathcal{T}$.
\end{enumerate}

In the basic implementation on the multi-core CPU platform the entire data required for the algorithm is processed in memory.

The GPU implementation of the basic algorithm uses the GPU's global memory to store all of the required data structures. Global memory on the GPU is large and reasonably efficient (if you carefully manage the memory access patterns) but considerably slower than much smaller shared and constant memories available on each streaming multi-processor.

\subsubsection{Optimised/Chunked Implementation}

The optimised version of the GPU algorithm endeavours to utilise shared and constant memory as much as possible. The key concept employed in the optimised algorithm is chunking. In this context, chunking means to process a block of events of fixed size (referred to as chunk size) for the efficient use of shared memory.

In the optimised GPU implementation, all of the three major steps (lines 3-19 in the basic algorithm, i.e., events in a trial and both financial and layer terms computations) of the basic algorithm are chunked. In addition, the financial terms, $\mathcal{I}$, and the layer terms, $\mathcal{T}$, are stored in the streaming multi-processor's constant memory. In the basic implementation, $l_{x_{d}}$ and $lo_{x_{d}}$ are represented in the global memory and therefore, in each step while applying the financial and layer terms the global memory has to be accessed and updated adding considerable overhead. This overhead is minimised in the optimised implementation.

\subsection{Results}
The results obtained from the basic and the optimised implementations are described in this section. The basic implementation results are presented for both multi-core CPU and many-core GPU platforms, while the optimised implementation is applicable only to the GPU.

\subsubsection{Results for Aggregate Analysis on CPUs}

\begin{figure*} [!tp]
	\subfloat[No. of Events in a Trial vs time taken for executing]{\label{fig:s11}\includegraphics[width=0.245\textwidth]{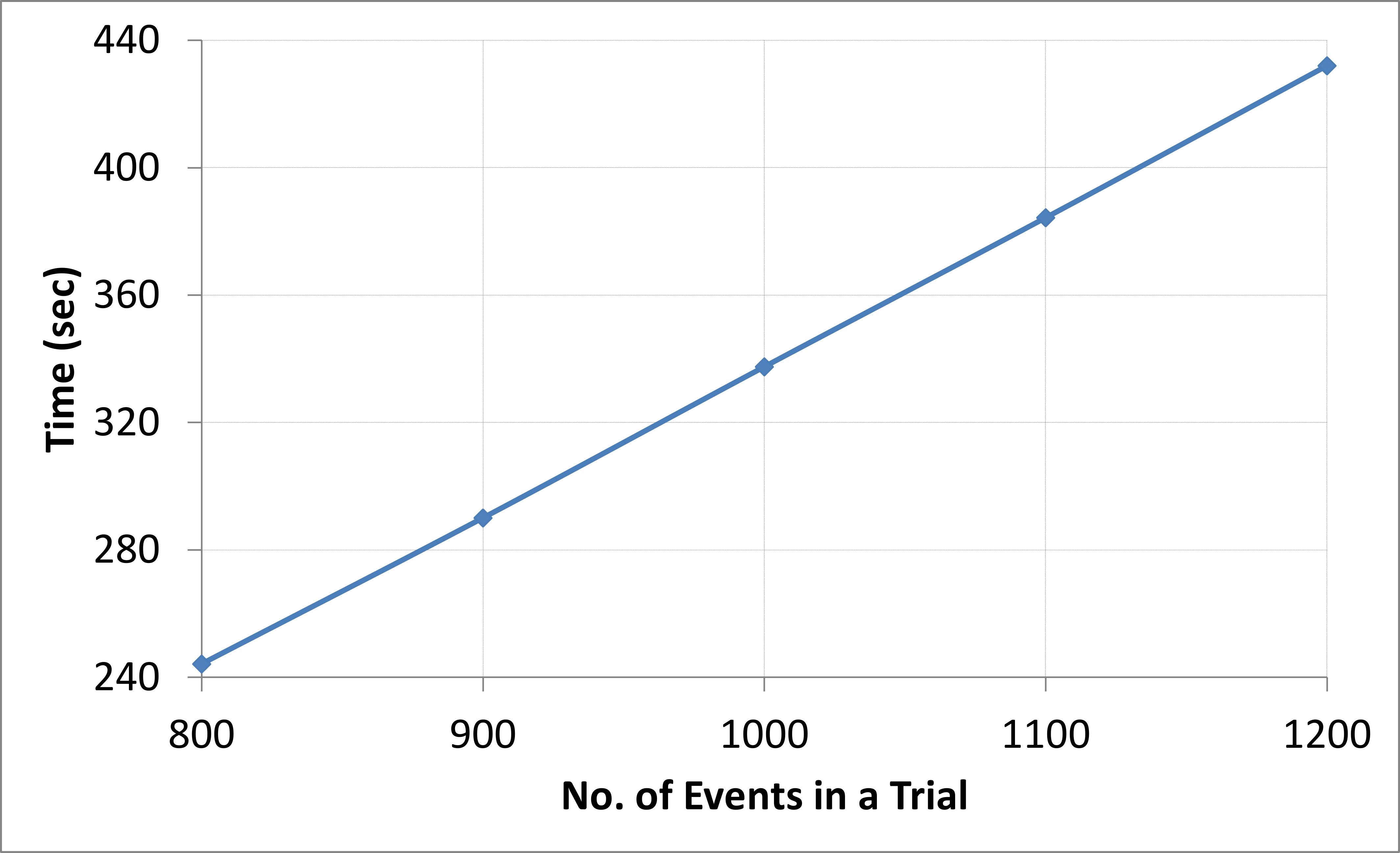}} \hfill
	\subfloat[No. of Trials vs time taken for executing]{\label{fig:s12}\includegraphics[width=0.245\textwidth]{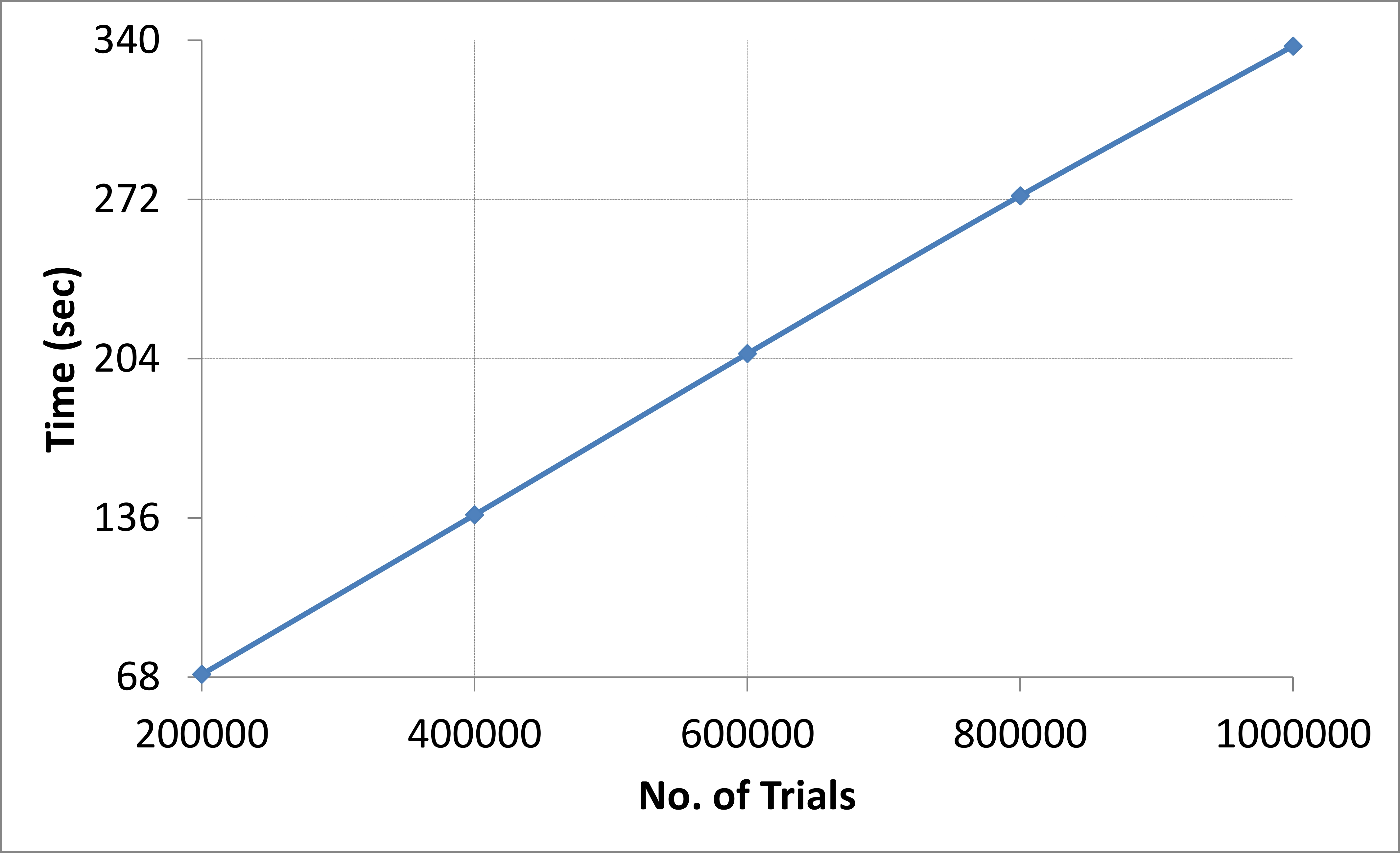}} \hfill
	\subfloat[Average number of ELTs per Layer vs time taken for executing]{\label{fig:s13}\includegraphics[width=0.245\textwidth]{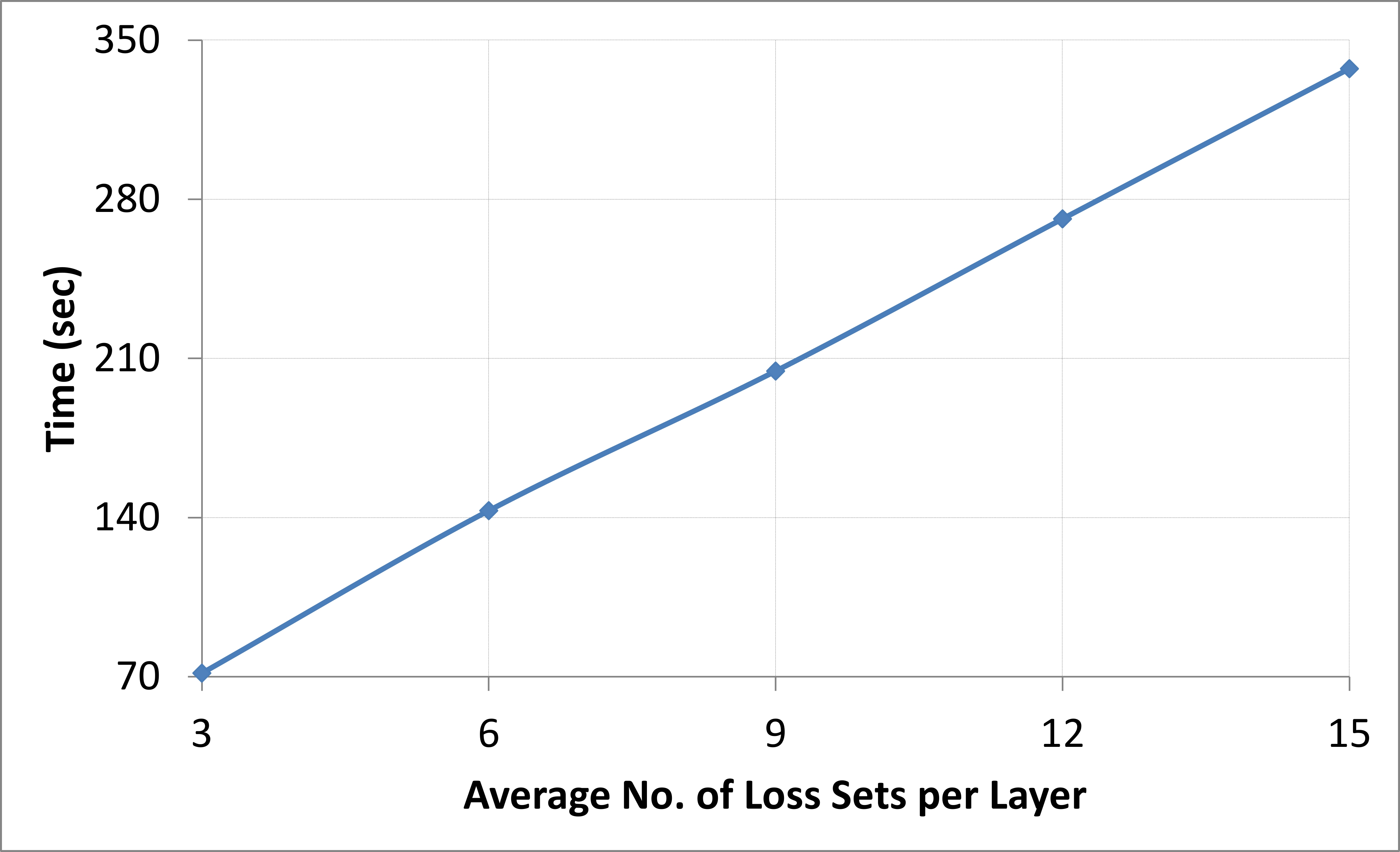}} \hfill
	\subfloat[Number of Layers vs time taken for executing]{\label{fig:s14}\includegraphics[width=0.245\textwidth]{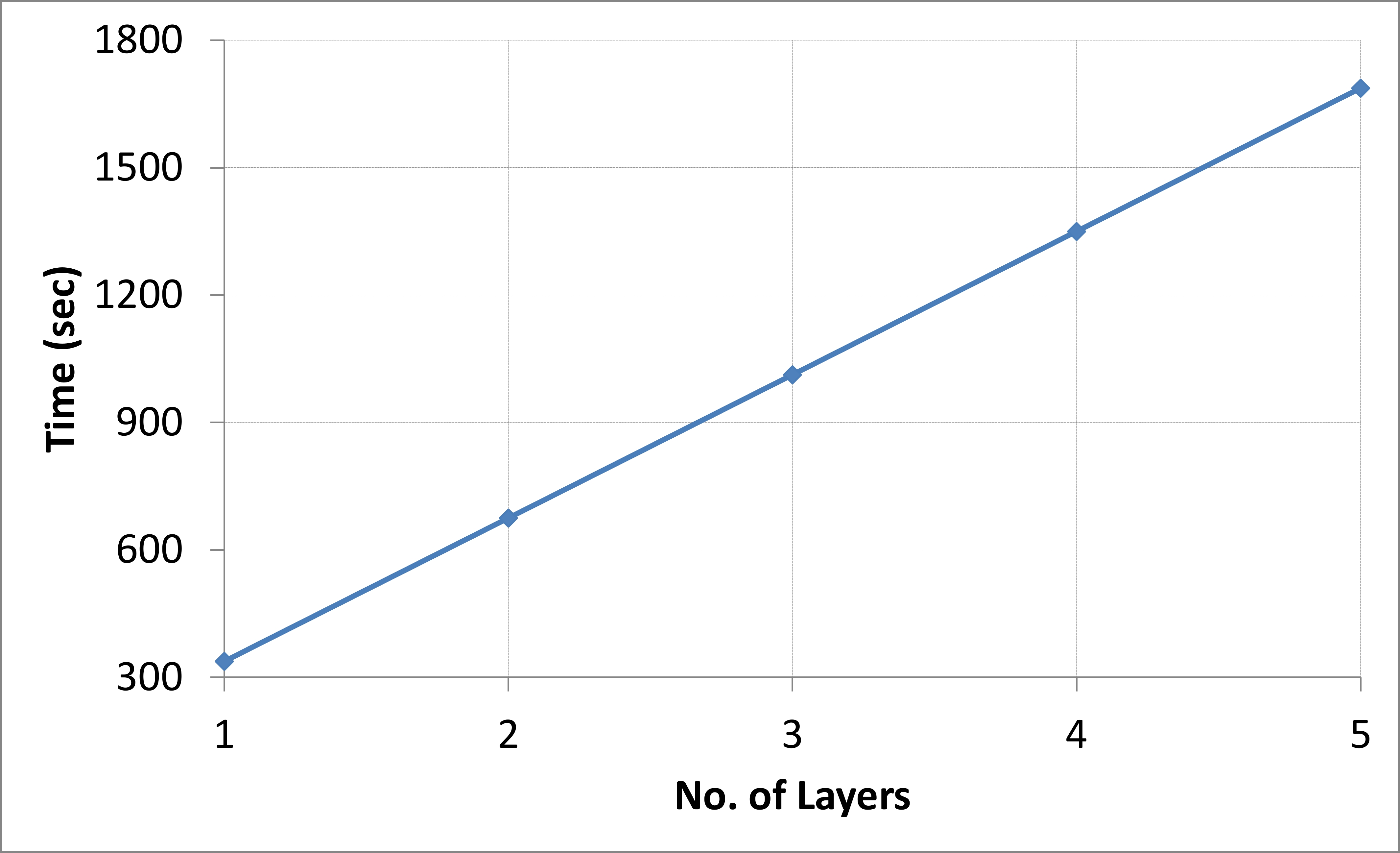}}\hfill
\caption{Performance of the basic aggregate analysis algorithm on a CPU using a single core}
\label{graphset1}
\end{figure*}

\begin{figure*} [!tp]
\centering
	\subfloat[No. of cores vs execution time]{\label{fig:s22}\includegraphics[width=0.495\textwidth]{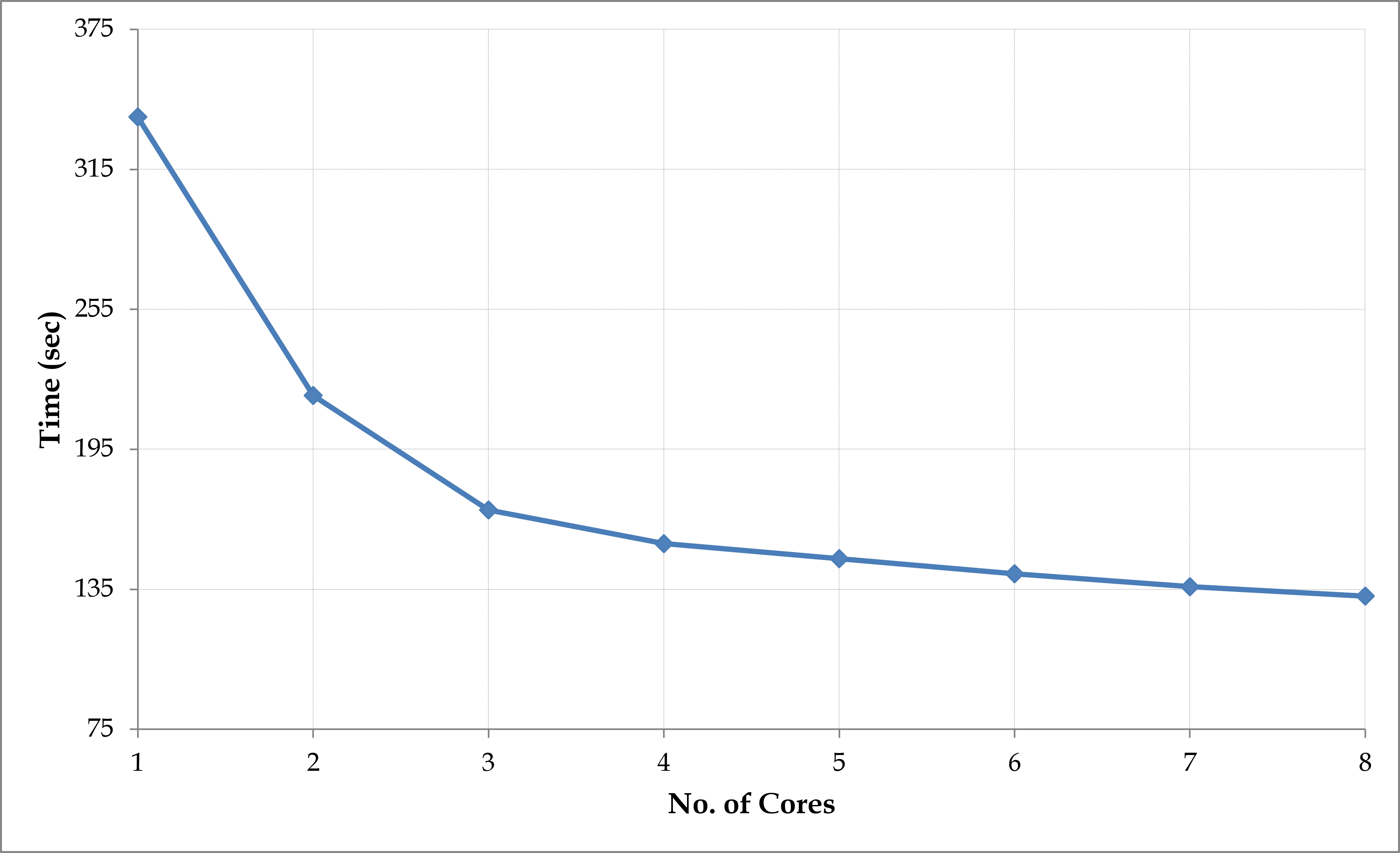}} \hfill
	\subfloat[Total No. of threads vs execution time]{\label{fig:s21}\includegraphics[width=0.495\textwidth]{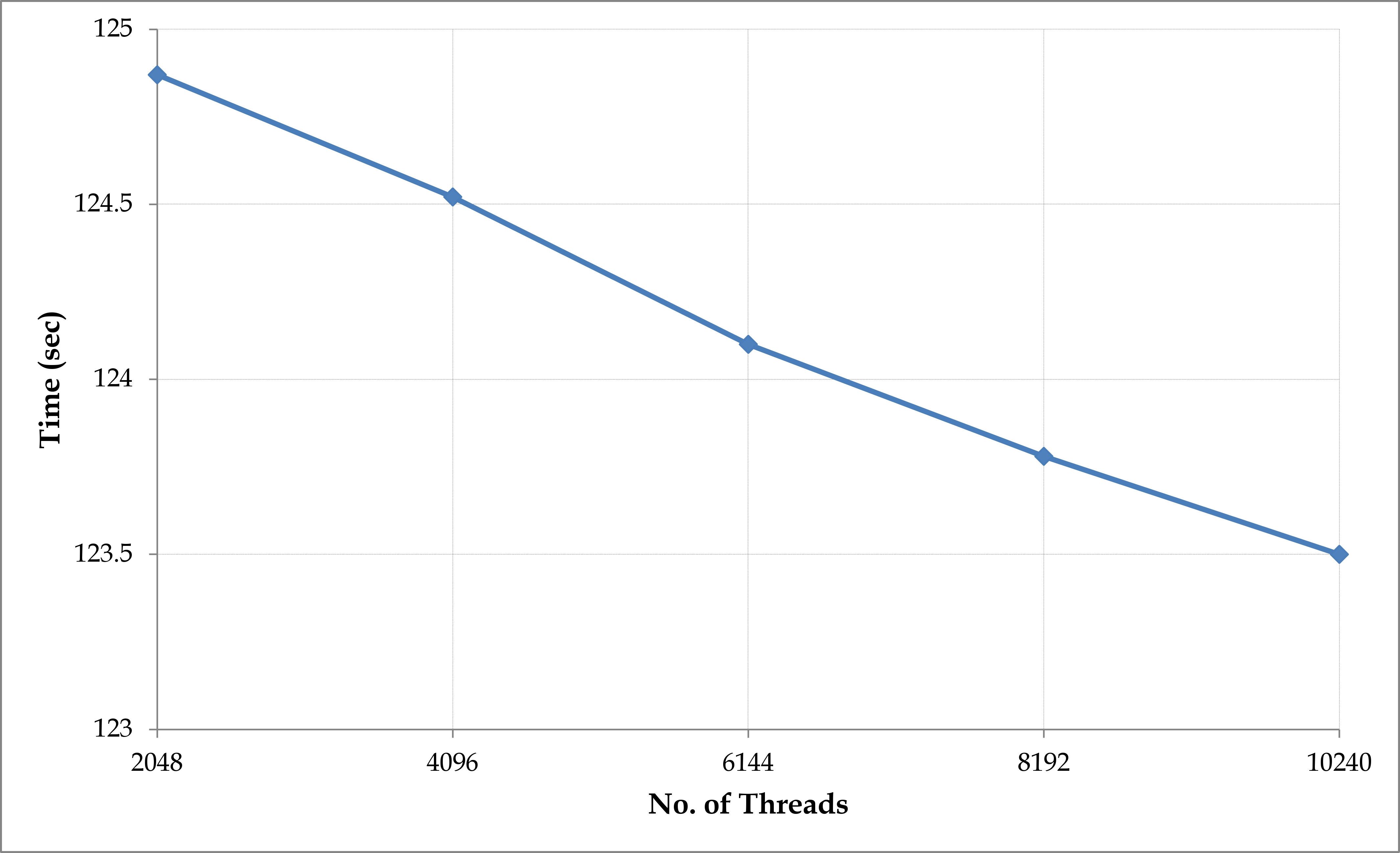}} \\
\caption{Graphs plotted for the parallel version of the basic aggregate analysis algorithm on a multi-core CPU}
\label{graphset2}
\end{figure*}

The size of an aggregate analysis problem is determined by four key parameters of the input, namely:
\begin{enumerate}
\item[(i)] Number of Events in a Trial, $|Et|_{av}$, which affects computations in line nos. 4-19 of the basic algorithm.
\item[(ii)] Number of Trials, $|T|$, which affects the loop in line no. 2 of the basic algorithm.
\item[(iii)] Average number of ELTs per Layer, $|ELT|_{av}$, which affects line no. 3 of the basic algorithm.
\item[(iv)] Number of Layers, $|L|$, which affects the loop in line no. 1 of the basic algorithm
\end{enumerate}

Figure \ref{graphset1} shows the impact on running time of executing the sequential version of the basic aggregate analysis algorithm on a CPU using a single core when the number of the number of events in a trial, number of trials, average number of ELTs per layer and number of layers is increased. The range chosen for each of the input parameters represents the range expected to be observed in practice and is based on discussions with industrial practitioners.

In Figure \ref{fig:s11} the number of ELTs per Layer is varied from 3 to 15. The number of Layers are 1, the number of Trials are set to 1 million and each Trial comprises 1000 events. In Figure \ref{fig:s12} the number of Trials is varied from 200,000 to 1,000,000 with each trial comprising 1000 events and the experiment is considered for one Layer and 15 ELTs. In Figure \ref{fig:s13} the number of Layers is varied from 1 to 5 and the experiment is considered for 15 ELTs per Layer, 1 million trials and each Trial comprises 1000 events. In Figure \ref{fig:s14} the number of Events in a Trial is varied from 800 to 1200 and the experiments is performed for 1 Layer, 15 ELTs per Layer and 100,000 trials.

Asymptotic analysis of the aggregate analysis algorithm suggests that performance should scale linearly in these parameters and this is indeed what is observed. In all the remaining performance experiments the focus is on a large fixed size input that is representative of the kind of problem size observed in practice.

Figure \ref{graphset2} illustrates the performance of the basic aggregate analysis engine on a multi-core CPU.  In Figure \ref{fig:s22}, a single thread is run on each core and the number of cores is varied from 1 to 8. Each thread performs aggregate analysis for a single trial and threading is implemented by introducing OpenMP directives into the C++ source. Limited speedup is observed. For two cores we achieve a speedup of 1.5x, for four cores the speedup is 2.2x, and for 8 cores it is only 2.6x. As we increase the number of cores we do not equally increase the bandwidth to memory which is the limiting factor. The algorithm spends most of its time performing random access reads into the ELT data structures. Since these accesses exhibit no locality of reference they are not aided by the processors cache hierarchy. A number of approaches were attempted, including the chunking method described later for GPUs, but were not successful in achieving a high speedup on our multi-core CPU. However a moderate reduction in absolute time by running many threads on each core was achieved.

Figure \ref{fig:s21} illustrates the performance of the basic aggregate analysis engine when all 8 cores are used and each core is allocated many threads. As the number of threads are increased an improvement in the performance is noted. With 256 threads per core (i.e. 2048 in total) the overall runtime drops from 135 seconds to 125 seconds. Beyond this point we observe diminishing returns as illustrated in Figure \ref{fig:s21}.

\subsubsection{Results for the basic aggregate analysis algorithm on GPU}

\begin{figure}[t]
\centering
	\includegraphics[width=0.5\textwidth]{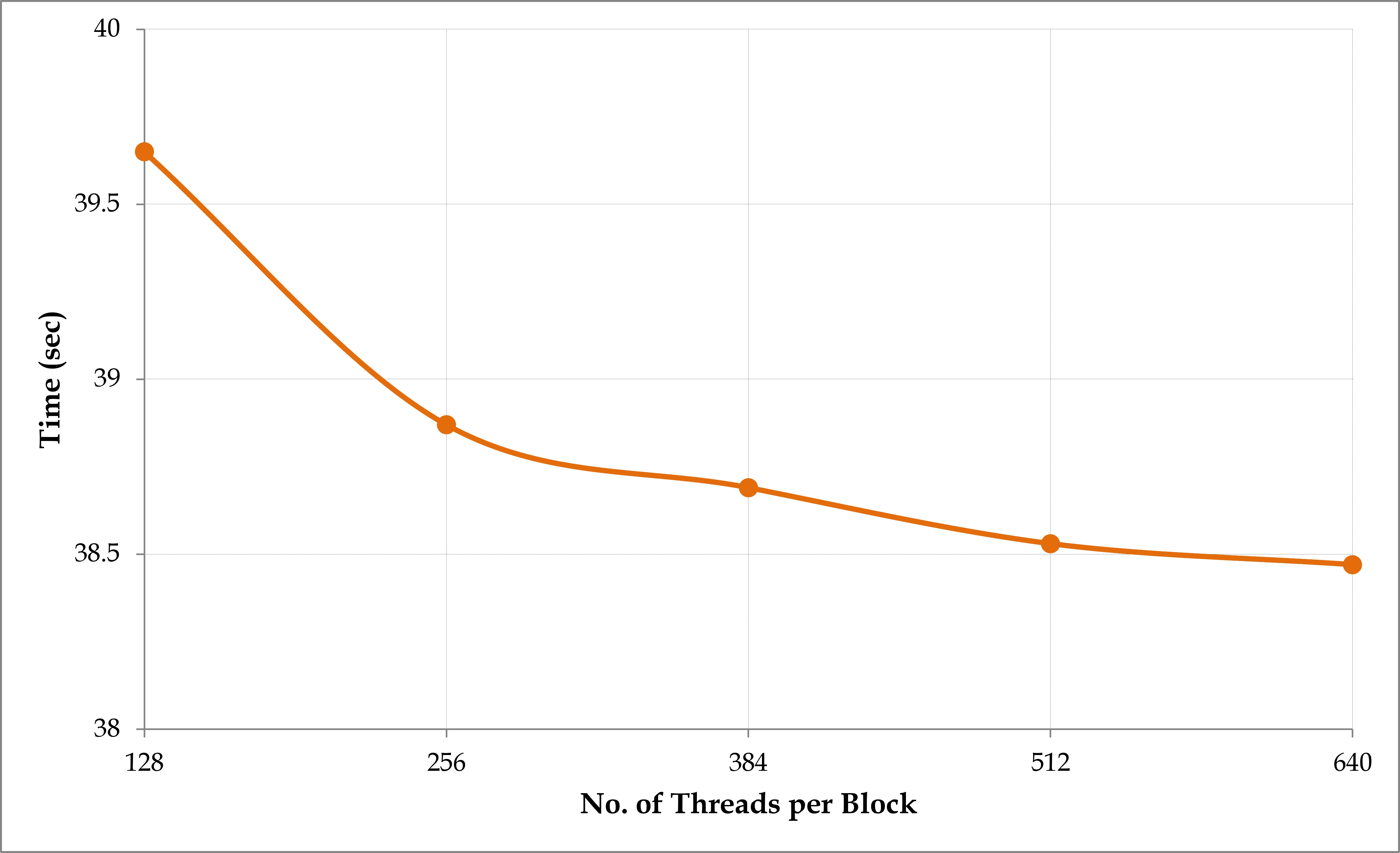}
	\caption{Graphs plotted for number of threads vs the time taken for executing the parallel version of the Basic Algorithm on many-core GPU}
	\label{graph3}
\end{figure}

In the GPU implementations, CUDA provides an abstraction over the streaming multi-processors, referred to as a CUDA block. When implementing the basic aggregate analysis algorithm on a GPU we need to select the number of threads executed per CUDA block.  For example, consider 1 million threads are used to represent the simulation of 1 million trials on the GPU, and 256 threads are executed on a streaming multi-processor. There will be $\frac{1,000,000}{256} \approx 3906$ blocks in total which will have to be executed on 14 streaming multi-processors. Each streaming multi-processor will therefore have to execute $\frac{3906}{14} \approx 279$ blocks. Since the threads on the same streaming multi-processor share fixed size allocations of shared and constant memory there is a real trade-off to be made. If we have a smaller number of threads, each thread can have a larger amount of shared and constant memory, but with a small number of threads we have less opportunity to hide the latency of accessing the global memory.

Figure \ref{graph3} shows the time taken for executing the parallel version of the basic implementation on the GPU when the number of threads per CUDA block are varied between 128 and 640. At least 128 treads per block are required to efficiently use the available hardware. An improved performance is observed with 256 threads per block but beyond that point the performance improvements diminish greatly.

\subsubsection{Results for the optimised aggregate analysis algorithm on GPU}

\begin{figure*}[t]
\centering
	\subfloat[Size of chunk vs time taken for executing]{\label{fig:s42}\includegraphics[width=0.495\textwidth]{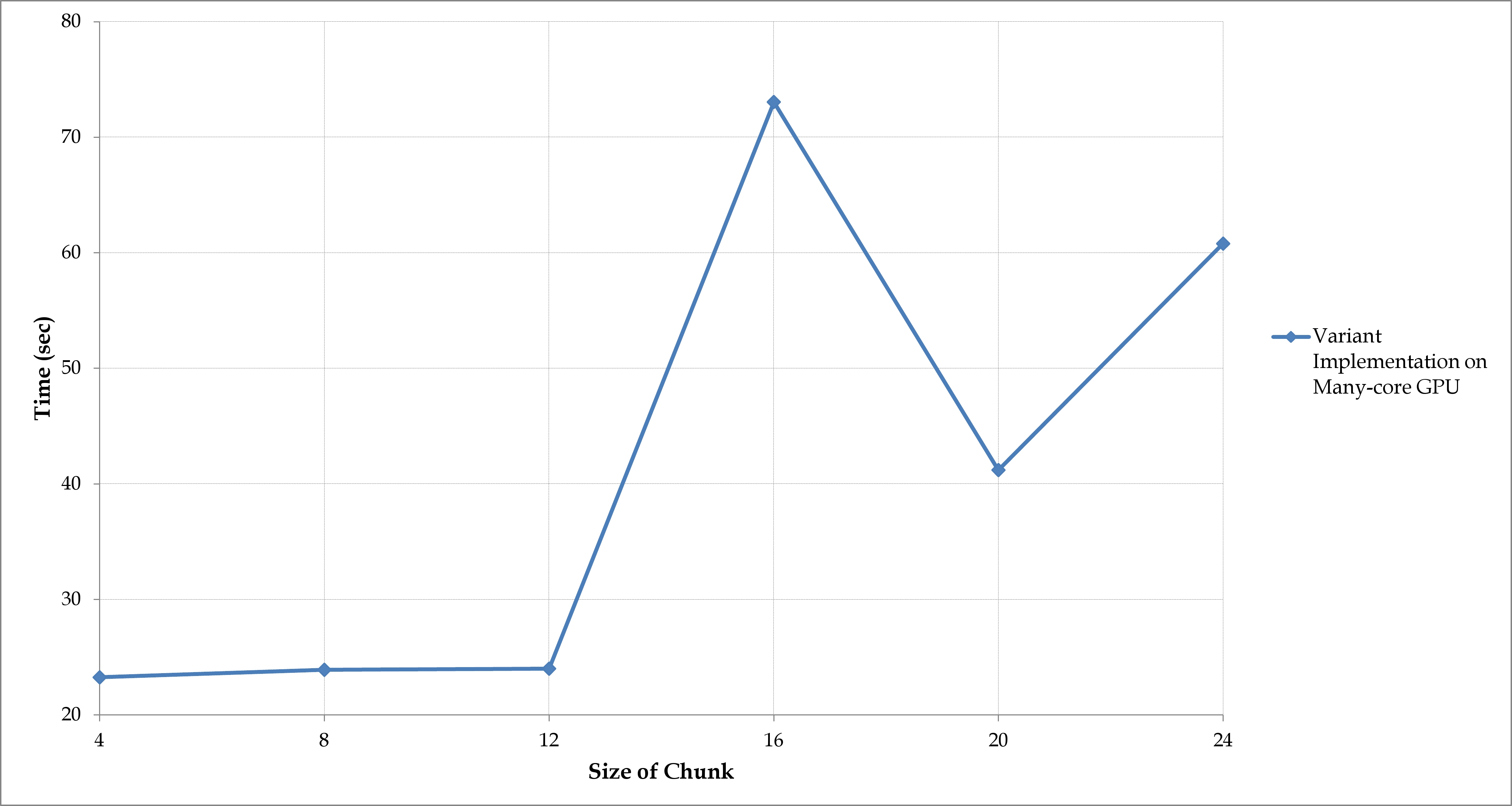}} \hfill
	\subfloat[No. of threads per block vs time taken for executing]{\label{fig:s41}\includegraphics[width=0.495\textwidth]{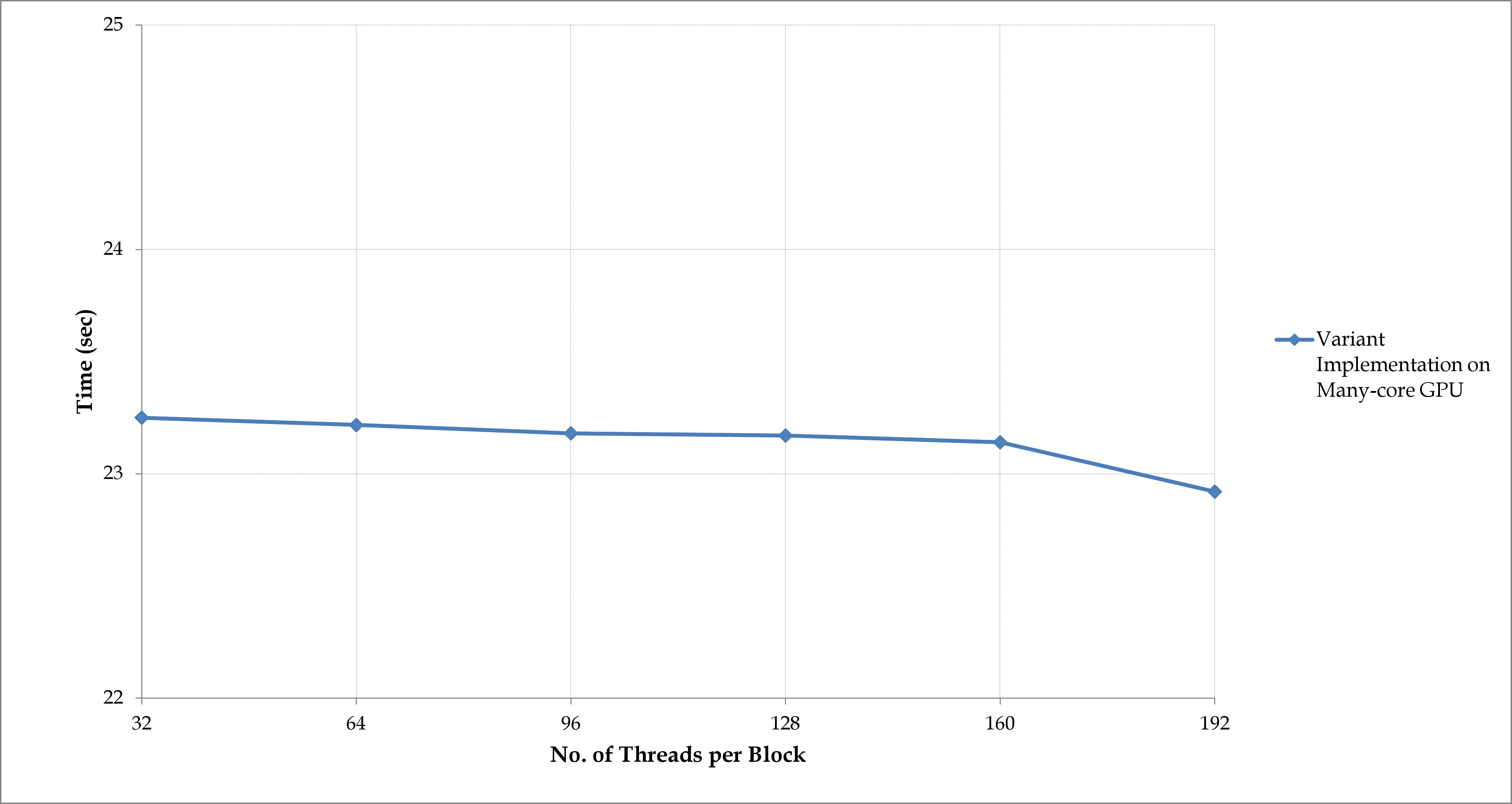}} \\
\caption{Performance of the optimised aggregate analysis algorithm on GPU}
\label{graphset4}
\end{figure*}

\begin{figure*}[t]
\centering
	\subfloat[Total time taken for executing the algorithm]{\label{fig:s61}\includegraphics[width=0.495\textwidth]{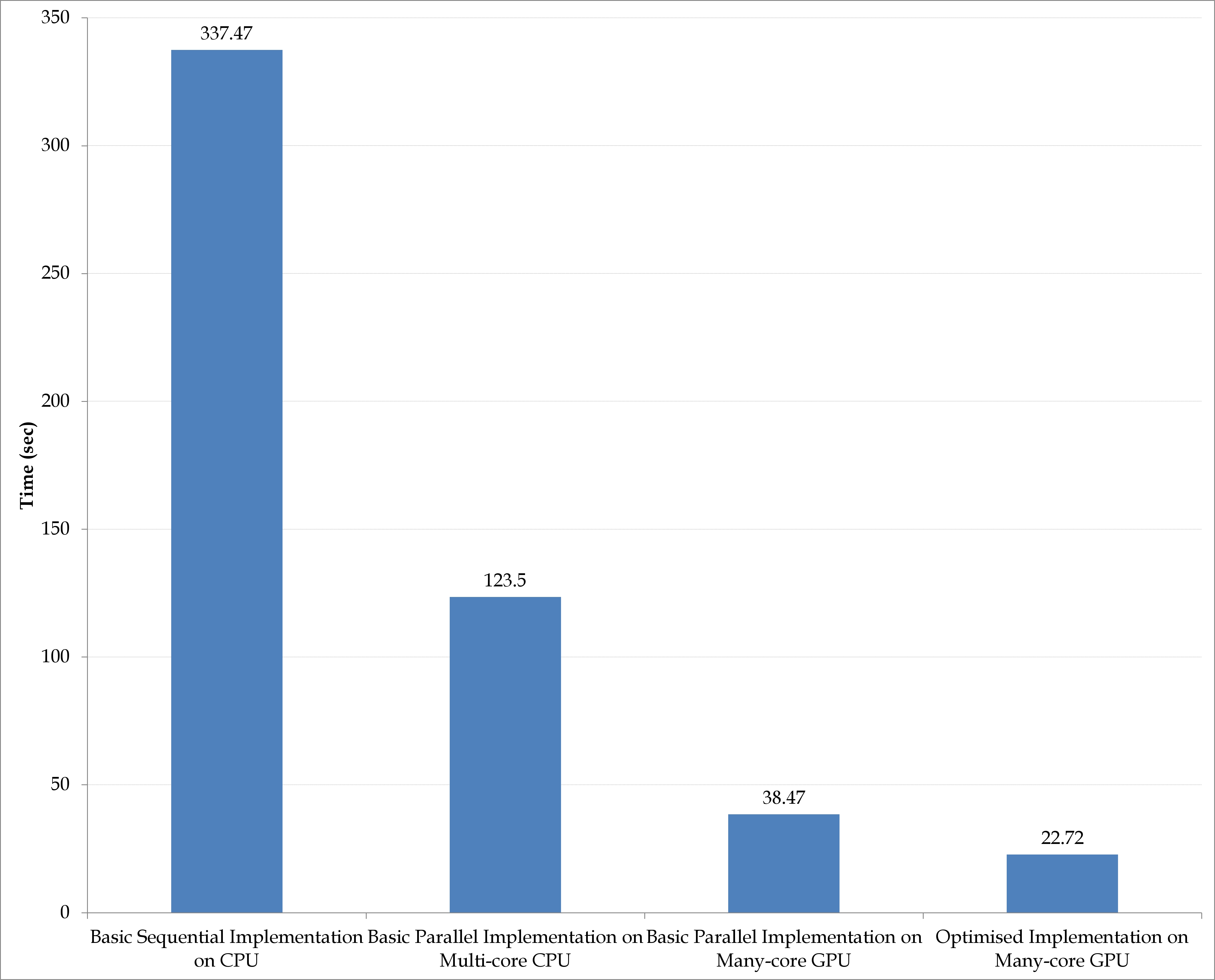}} \hfill
	\subfloat[Percentage of time taken for fetching Events from memory, time for look-up of ELTs in the direct access table, time for financial term calculations and time for layer term calculations]{\label{fig:s62}\includegraphics[width=0.495\textwidth]{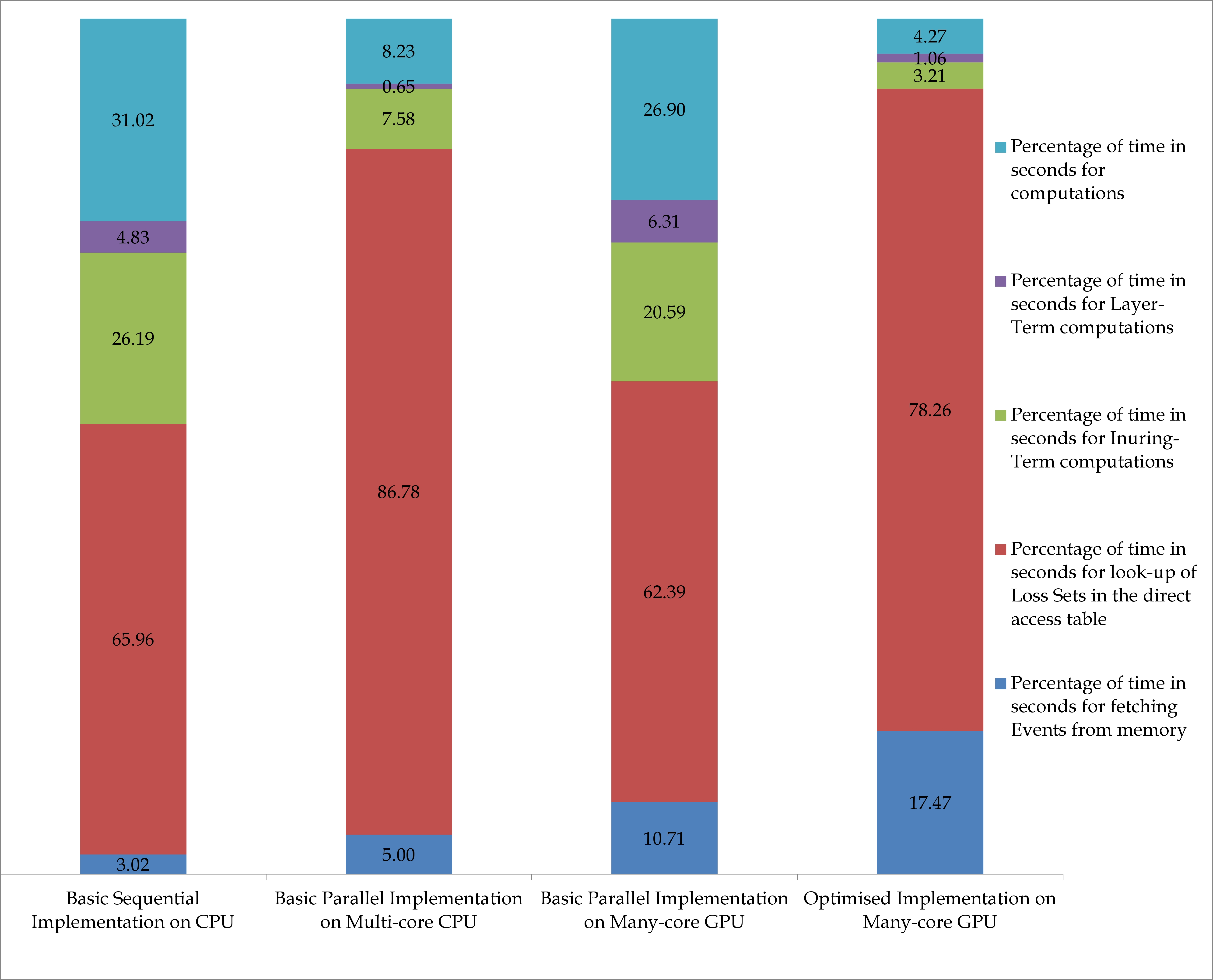}}
	\caption{Summary of results of the Basic and Optimised Implementations}
	\label{graph6}
\end{figure*}

The optimised or chunked version of the GPU algorithm aims to utilise shared and constant memory as much as possible by processing ``chunks'', blocks of events of fixed size (referred to as chunk size), to improve the utilisation of the faster shared memories that exist on each streaming multi-processor.

Figure \ref{fig:s42} illustrates the performance of the optimised aggregate analysis algorithm as the chunk size is increased. With a chunk size of 4 the optimised algorithm has a significantly reduced runtime from 38.47 seconds down to 22.72 seconds, representing a 1.7x improvement. Interestingly, increasing the chunk size does not improve the performance. The curve is observed to be flat up to a chunk size of 12 and beyond that the performance deteriorates rapidly as the shared memory overflow is handled by the slow global memory.

Figure \ref{fig:s41} illustrates the performance of the optimised aggregate analysis algorithm as the number of threads is increased. The number of threads range in multiples of 32 due to the WARP size, which corresponds to the number of symmetric multi-processors in CUDA. With a chunk size of 4 the maximum number of threads that can be supported is 192. As the number of threads are increased, there is a small gradual improvement in performance, but the overall improvement is not significant.

\section{Discussion and Conclusion}
\label{conclusion}
The performance results for the various algorithms described in this paper are summarised in Figure \ref{graph6}. Each bar in Figure \ref{fig:s61} represents the total time taken for executing each algorithm with 1 million trial, each trial consisting of 1000 events. In each case, the algorithm specific tuning parameters (such as the number of threads) were set to the best value identified during experimentation. The performance on the Intel i7 multi-core hardware is disappointing both in terms of absolute performance and speedup. It is not clear how to improve performance or speedup as the bottleneck is memory-bandwidth.

Figure \ref{fig:s62} breaks down the total percentage of time taken for the main phases of the algorithm, namely, (a) time for fetching events from the memory, (b) time for look-up of ELTs in the direct access table, (c) time for financial term calculations and (d) time for layer term calculations. 78\% of the time is taken for accessing the ELT data structures in memory. These accesses have been optimised by implementing the data structure as a direct access table. Given this bottleneck there does not appear to be much room for significant algorithmic performance improvements. If selecting alternative hardware is feasible, use of multi-core processors with a higher memory bandwidth may improve speedup. The current financial calculations can be implemented using basic arithmetic operations. However, if the system is extended to represent losses as a distribution (rather than a simple mean) then the algorithm would likely benefit from use of a numerical library for convolution.

The performance of the GPU implementations is quite promising as they are significantly faster than on their multi-core counterparts. The basic GPU method is 3.2x faster, while the optimised version is 5.4x faster. In absolute terms the optimised GPU algorithm can perform a 1 million trial aggregate simulation on a typical contract in just over 20 seconds. This is sufficiently fast to support a real-time pricing scenario in which an underwriter can evaluate different contractual terms and pricing while discussing a deal with a client over the phone. In many applications 50K trials may be sufficient in which case sub one second response time can be achieved. Aggregate analysis using 50K trials on complete portfolios consisting of 5000 contracts can be completed in around 24 hours which may be sufficiently fast to support weekly portfolio updates performed to account for changes such as currency fluctuations. If a complete portfolio analysis is required on a 1M trial basis then a multi-GPU hardware platform would likely be required.

To conclude, the research reported in this paper investigates parallel methods for aggregate risk analysis which utilises large input data, organise input data in efficient data structures, and define the granularity at which parallelism can be applied on the aggregate risk analysis problem to achieve a significant speedup. To this end, a parallel aggregate risk analysis algorithm and an engine based on the algorithm is proposed. The engine is implemented on both multi-core CPU and GPU platforms on which experimental studies are pursued. The results obtained on the GPU confirm the feasibility of performing fast aggregate analysis using large data in real-time on relatively low cost hardware compared to the costly large-scale clusters that are currently being employed.

\end{document}